\documentclass[preprint]{aastex6}
\pdfoutput=1

\usepackage{amsmath}
\usepackage{selectp}
\usepackage{mathtools,amssymb}
\usepackage{multirow}
\usepackage[utf8]{inputenc}
\usepackage{nicefrac}
\usepackage{xfrac}
\usepackage{graphicx}
\usepackage{comment}

\newcommand{\til}{\raise.35ex\hbox{$\scriptstyle\sim$}}

\begin{document}

\title{The Short Rotation Period of Hi'iaka,\\ Haumea's Largest Satellite}
\author{Danielle M. Hastings\altaffilmark{1,2}, Darin Ragozzine\altaffilmark{2,3}, Daniel C. Fabrycky\altaffilmark{4}, Luke D. Burkhart\altaffilmark{5,6}, Cesar Fuentes\altaffilmark{7}, Jean-Luc Margot\altaffilmark{1}, Michael E. Brown\altaffilmark{8}, Matthew Holman\altaffilmark{5}}
\email{dhastings@g.ucla.edu}

\altaffiltext{1}{University of California, Los Angeles, Department of Earth, Planetary, and Space Sciences, 595 Charles Young Drive East, Los Angeles, CA 90095, USA}
\altaffiltext{2}{Florida Institute of Technology, Department of Physics and Space Sciences, 150 West University Boulevard, Melbourne, FL 32901, USA}
\altaffiltext{3}{Brigham Young University, BYU Department of Physics and Astronomy N283 ESC, Provo, UT 84602, USA}
\altaffiltext{4}{Department of Astronomy and Astrophysics, University of Chicago, 5640 South Ellis Avenue, Chicago, IL 60637, USA}
\altaffiltext{5}{Harvard-Smithsonian Center for Astrophysics, 60 Garden Street, Cambridge, MA 02138, USA}
\altaffiltext{6}{Yale University, Department of Physics, 217 Prospect St, New Haven, CT 06511, USA}
\altaffiltext{7}{Departamento de Astronomía, Universidad de Chile, Camino El Observatorio 1515, Santiago, Chile}
\altaffiltext{8}{California Institute of Technology, Division of Geological and Planetary Sciences, MC 150-21, Pasadena, CA 91125, USA}

\slugcomment{Rapid Rotation of Hi'iaka}

\begin{abstract}

Hi'iaka is the larger outer satellite of the dwarf planet Haumea. Using relative photometry from the Hubble Space Telescope and Magellan and a phase dispersion minimization analysis, we have identified the rotation period of Hi'iaka to be $\sim 9.8$ hrs (double-peaked).  This is $\sim 120$ times faster than its orbital period, creating new questions about the formation of this system and possible tidal evolution.  The rapid rotation suggests that Hi'iaka could have a significant obliquity and spin precession that could be visible in light curves within a few years. We then turn to an investigation of what we learn about the (presently unclear) formation of the Haumea system and family based on this unexpectedly rapid rotation rate. We explore the importance of the initial semi-major axis and rotation period in tidal evolution theory and find they strongly influence the time required to despin to synchronous rotation, relevant to understanding a wide variety of satellite and binary systems. We find that despinning tides do not necessarily lead to synchronous spin periods for Hi'iaka, even if it formed near the Roche limit.  Therefore the short rotation period of Hi'iaka does not rule out significant tidal evolution.  Hi'iaka's spin period is also consistent with formation near its current location and spin up due to Haumea-centric impactors.  

\end{abstract}

\keywords{Kuiper belt objects: individual (Haumea) --- planets and satellites: dynamical evolution and stability --- planets and satellites: individual (Hi'iaka) --- techniques: photometric}

\maketitle

\section{Introduction} \label{introduction}

The dwarf planet Haumea stands out from the rest of its Kuiper Belt counterparts.  It has the shortest known rotation period ($P_{rot}=3.9154$ hrs) of objects its size \citep{2006ApJ...639.1238R}.  Haumea is also known to have two regular satellites, Hi'iaka and Namaka \citep{2009AJ....137.4766R}, and a collisional family of smaller objects associated with it \citep{2434964720070315}. These family members share many unusual properties with Haumea including strong water ice spectra \citep[][]{2008ApJ...684L.107S,2011ApJ...730..105T,2012A&A...544A.137C}, high albedos \citep{2010Natur.465..897E}, and possibly a more rapid mean rotational period of $6.27 \pm 1.19$ hrs compared to a mean rotational period of $7.65 \pm 0.54$ hrs \citep{2016arXiv160304406T} for other Kuiper Belt objects (KBOs), in addition to their dynamically clustered orbits \citep{2011ApJ...733...40M,2012MNRAS.421.1331L,2012Icar..221..106V}. All of these properties point to formation by a major collision which can impart rapid spin to Haumea and generate the satellites and family.

Existing formation hypotheses cannot self-consistently explain all the properties of Haumea's formation \citep{2012MNRAS.419.2315O,Campo-Bagatin13062016}. For example, models which invoke a slow impactor to keep the Haumea family very tightly clustered \citep{2010ApJ...714.1789L} are improbable \citep{2008AJ....136.1079L,Campo-Bagatin13062016}. These mechanisms might be reconciled if Haumea and possibly other large KBOs were near-equal size binaries that were eventually destabilized, potentially due to three-body dynamical effects of the Sun \citep{2011ApJ...733...40M,2012Icar..220..947P,2012AJ....143..146B}. Though this can create a slow impactor without relying on a low-probability heliocentric impact, whether it is more plausible than other hypotheses requires further study. 

One avenue for improving our understanding of the formation of Haumea is to study its two moons. Their nearly circular and coplanar orbits suggest that they formed as a direct consequence of the same event that spun up Haumea (though it is not impossible that this was a different event from the formation of the family \citep{2009ApJ...700.1242S}). Therefore, their physical and orbital properties may contain important clues. 

Hi'iaka and Namaka have nominal masses of $\sim$0.5\% and $\sim$0.05\% of Haumea's mass ($M_H$), where $M_H=(4.006\pm 0.040)\times10^{21}$ kg, and nominal radii of 150 and 75 km, respectively \citep{2009AJ....137.4766R}.  Due to uncertainties in density and brightness measurements, these values may have uncertainties on the level of tens of percent. The satellites have large semi-major axes, orbiting at 35.7 and 69.5 Haumea radii ($R_H$), where we use the volumetric Haumea radius of $R_H \simeq 715$ km (from 495 x 770 x 960 km estimated in \citet{Lockwood2014}).  The smaller inner satellite, Namaka, orbits with an eccentricity of 0.2 and an inclination of $13^{\circ}$.  The larger outer satellite, Hi'iaka, has a less excited orbit, with an eccentricity of 0.05 and an inclination of $\sim$2$^{\circ}$ \citep{2013AJ....146...89C}. A deep search essentially ruled out additional regular satellites as small as $\sim$10$^{-6}$ of Haumea's mass \citep{2016AJ....151..162B}. 

The larger satellite, Hi'iaka, orbits with a period of 49.462 $\pm$ 0.083 days \citep{2009AJ....137.4766R}.  At this point in Hi'iaka's orbital evolution, it is expected to have been tidally despun and therefore rotating synchronously (or potentially in a higher-order spin-orbit resonance) with its orbital period. Assuming standard tidal theory, the large semi-major axis and low eccentricity of Hi'iaka would take much longer to achieve than despinning of a small satellite. In many tidal histories, the despinning of a small satellite is often considered to be effectively instantaneous due to the short timescales involved. 

We present observations of Hi'iaka that clearly show that it is rotating $\sim$120 times faster than the expected synchronous spin period (Sections \ref{data} and \ref{analysis}). Such a configuration is unusual for a regular satellite, as other regular satellites in the solar system are tidally despun, although it mirrors the recent discovery of rapidly rotating small moons of Pluto \citep{2016Sci...351.0030W}. We then consider the implications for this rapid rotation in Section \ref{hypotheses} by considering two end-member possibilities for Hi'iaka's formation: formation near the Roche limit of Haumea in a standard post-impact disk (Section \ref{close_in}) or formation near the present-day location (Section \ref{far_out}). This instigates an extensive discussion on the validity of using standard tidal ``timescales'' which suggests that initial conditions are very important, even for extensive tidal evolution, as demonstrated by numerical integration.  We also consider the case of spin up by a recent impactor (Section \ref{spin_up}). We then draw conclusions and suggest future investigations in Section \ref{conclusion}. 

\section{Data}
\label{data}

As the goal is to identify Hi'iaka's light curve shape and period, only relative photometry is required. This simplifies the analysis considerably since our observations come from different telescopes under different observing conditions. Our primary data comes from Hubble Space Telescope (HST) observations on February 4, 2009 and June 28, 2010 and Magellan observations on June 1, 2009. 

The HST observations of the Haumea system comprised 5 HST orbits worth of 100-second exposures of the Wide Field Planetary Camera 2 (Program 11971) and 10 HST orbits worth of 44-second exposures of the Wide Field Camera 3 (Program 12243). Hereafter, we will refer to these as the "2009" and "2010" HST data, respectively. The 2009 observations were collected in an attempt to observe a possible mutual event between Hi'iaka and Namaka. They were well separated from Haumea, whose PSF was removed. These satellites were too close to resolve at this epoch, so simple aperture photometry with a 4-pixel radius circular aperture was used to determine the light curve. We return to the implications for Namaka later, but for now assume that the light curve is due entirely to Hi'iaka. 

\begin{figure}
    \centering
    \includegraphics[scale=.75]{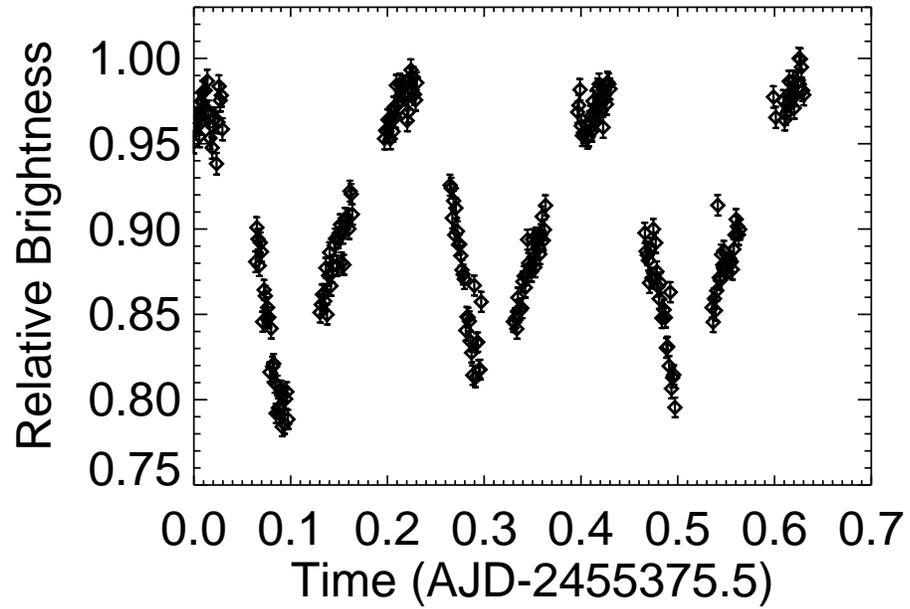}
    \caption{Unphased light curve for the 2010 HST data showing a strong repeated variability. These data only have Hi'iaka in the aperture. Investigations of other regions at the same Haumea-centric distance show no sign of variability. The error bars are estimated from photon noise.  Gaps in the light curve are due to Earth occultations.}
	\label{figure:lightcurve}
\end{figure}

For the 2010 HST data, all three objects are resolved for the first four orbits. Triple PSF-fits were performed as described in \citet{2009AJ....137.4766R} and \citet{2016AJ....151..162B} to identify the exact locations of Haumea, Hi'iaka, and Namaka and to remove Haumea's PSF from the images. In the last six orbits, Namaka is too close to Haumea to resolve (these observations were chosen to capture a Haumea-Namaka mutual event, so this is unsurprising), and double PSF-fits are used. In either case, Hi'iaka was far from Haumea and easily resolved. Simple aperture photometry was collected using a 4-pixel radius circular aperture, which is sufficient for our purposes. The PSF fits were designed for astrometry and do not return as reliable photometry. For more details on these observations see \citet{2016AJ....151..162B}. Several tests confirmed that the variability was real and centered on Hi'iaka. For example, investigation of the light curve of an identical aperture located opposite to Hi'iaka showed no significant variability. A few observations were significant outliers (due to cosmic ray hits) and were removed from our data; contamination from cosmic rays was also the primary motivation of the choice of aperture size. Gaps due to Earth occultations are a larger concern, but do not preclude the 2010 data from showing a strong repeated variability which we illustrate in Figure \ref{figure:lightcurve}.

This system was also observed on the night of UT June 2, 2009 with the Baade Magellan telescope at Las Campanas Observatory in Chile. We used the Raymond and Beverly Sackler Magellan Instant Camera (MagIC).  Observations were taken from the beginning of the night until it was unobservable, for a total of $\sim$5 hours. We centered the system on one of the four quadrants defined by the instrument's four amplifiers. The seeing was constant during the observations and consistently close to $0.5 ''$, smaller than Hi'iaka's separation of $1.4 ''$. The  SITe CCD detector has a pixel scale of 0.069 arcsec per pixel. We set the exposure times at 120 seconds to avoid saturation and optimize readout time. The filter selected was Johnson-Cousins R. Standard calibrations were taken at the beginning and end of the night. The telescope guiding system ensured the pointing was constant to within a FWHM over the course of the observations.

Standard routines were used to trim, bias-subtract, and flat-field the images. Each exposure was then registered using the ISIS package \citep{2000A&AS..144..363A} to PSF match and subtract a template. The template image was the combination of the twenty sharpest images, using an average with sigma rejection on each pixel. Ordinary aperture photometry was then applied on the subtracted images to Haumea and two comparison stars of similar brightness in the field of view using the DAOPHOT II package \citep{1538-3873-99-613-191, 1992ASPC...25..297S}. To remove the influence of Haumea on photometry at Hi'iaka's location, a two-dimensional Gaussian that was the best-fit to Haumea was subtracted from the images.

For all of our observations, we can be confident that Hi'iaka's photometry was not affected by Haumea's variability because we see no sign of Haumea's large amplitude ($\sim$25\%) 3.9-hour rotational light curve. While it is possible that very minor contamination remains, it is far exceeded by the highly significant variations in Hi'iaka's light curve and does not affect our conclusions.

Each dataset has been normalized to the respective maximum Hi'iaka brightness (all three go through a maximum), in order to provide relative photometry. We also investigated HST data from 2008 (Program 11518) and 2014-15 (Program 13873) with this rotation period. These data are composed of single-orbit investigations separated by weeks, have larger systematic errors, multiple filters, and much lower cadence ($\sim$15 minutes). However, they showed the same types of trends as seen in the higher cadence light curves: within an orbit, Hi'iaka's brightness could change by roughly $\pm$10\%. Other datasets are thus consistent with our conclusions. 

Due to the different observation times, the Hi'iaka-observatory distance changes significantly, introducing light-travel time variations. Therefore, all times are converted to ``HaumeA-centered Julian Date'' (AJD), a clock local to the Haumea system and therefore mutually self-consistent. Table \ref{table:data} presents these relative normalized photometry inferred from our observations.

\begin{deluxetable}{cccc}

\tablecolumns{3}
\tablewidth{0pt}
\tablecaption{Normalized Relative Photometry of Haumea\label{table:data}}
\tablehead{\colhead{AJD} & \colhead{Normalized Flux} & \colhead{Normalized Errors} & \colhead{Observing Program}\\
\colhead{(d)} & \colhead{} & \colhead{} & \colhead{}} 
\startdata
2454867.136 & 0.9898 & 0.0103 & 1\\
2454867.138	& 0.9797 & 0.0102 & 1\\
2454867.157 & 0.9101 & 0.0099 & 1\\
2454867.159 & 0.9124 & 0.0099 & 1\\
2454867.161	& 0.9222 & 0.0099 & 1\\
2454867.163 & 0.9136 & 0.0099 & 1\\
2454867.203	& 0.8352 & 0.0095 & 1\\
2454867.205	& 0.8360 & 0.0095 & 1\\
2454867.207	& 0.8253 & 0.0094 & 1\\
2454867.209	& 0.8397 & 0.0095 & 1\\
\enddata
\tablecomments{Table \ref{table:data} is published in its entirety in the machine-readable format.  A portion is shown here for guidance regarding its form and content.
AJD is the HaumeA-centered Julian Date of the observations, after light travel time corrections. The normalized flux and errors are derived from relative photometry measurements from HST Program 11971 on February 4, 2009 (1), Magellan observations on June 1, 2009 (2), and HST Program 12243 on June 28, 2010 (3). The photometry was normalized by dividing each individual dataset by the maximum value of flux from that dataset.}

\end{deluxetable}

\section{Analysis and Results}
\label{analysis}

\subsection{Period Analysis}
\label{period}

The raw HST relative photometry given in Table \ref{table:data} showed an extremely significant variability with a periodicity near 5 hours and all three datasets indicated a similar sawtoothed shape. To identify a specific period, we employed phase dispersion minimization (PDM) using the IDL routine \texttt{PDM2} (Marc Buie, personal communication). PDM typically involves minimizing the dispersion of the data at a given phase \citep{1978ApJ...224..953S}, but PDM2 seeks instead to minimize the reduced $\chi^2$ statistic in order to determine the best period \citep{1992Icar..100..288B}. We searched periods from 2 to 20 hours to find a period that produced a self-consistent phased light curve. We note that the different observation geometries due to the heliocentric motion of Haumea and Earth (and any precession of Hi'iaka) only span $\sim$5$^{\circ}$ in viewing angle, so secular changes in Hi'iaka's light curve would be minimal and PDM remains an appropriate technique. 

\begin{figure}
    \centering
    \includegraphics[scale=.75]{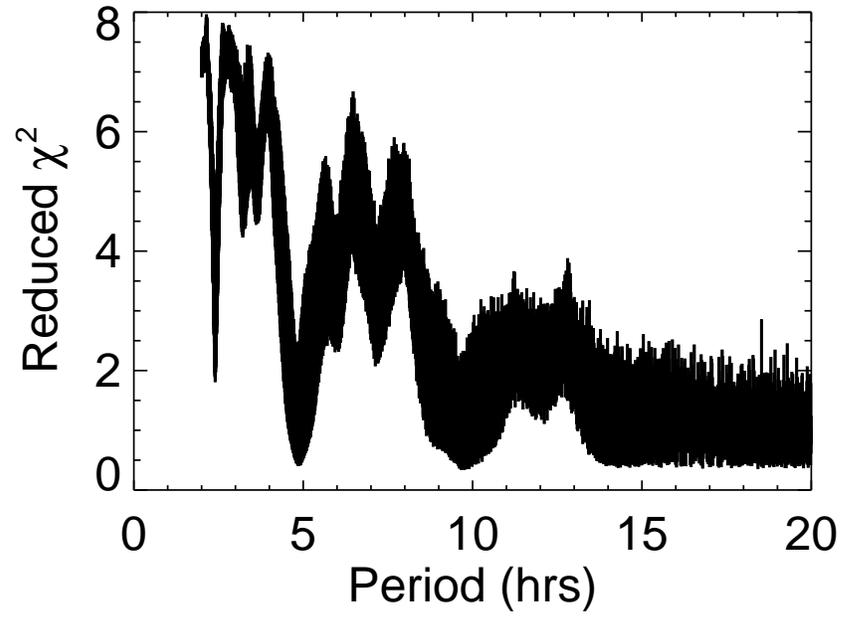}
    \caption{PDM periodogram for the Hi'iaka light curve data.  The minimum reduced $\chi^2$ values correspond to the most likely rotation periods for Hi'iaka.  The two regions of minima at 4.9 and 9.8 hours correspond to the single-peaked and double-peaked light curves, respectively.  The double-peaked period has the lowest reduced $\chi^2$ value and is preferred because Hi'iaka's significant variability is most likely due to the variable projected cross-sectional area of a rotating non-spherical body.}
	\label{figure:pdm}
\end{figure}

The resulting periodogram from PDM2 is shown in Figure \ref{figure:pdm}.  There are clearly two regions that are favored, trial periods of $\sim4.9$ and $\sim9.8$ hours, which correspond to the single-peaked and double-peaked light curves, respectively. The rotation period would correspond to the single-peaked light curve if it were caused by albedo variegations, but this is atypical for objects the size of Hi'iaka. We therefore identify the 9.8-hour period as the rotation period of Hi'iaka, with the double-peaked light curve resulting from variable projected cross-sectional area of a rotating non-spherical body.

\begin{figure}
    \centering
    \includegraphics[scale=.75]{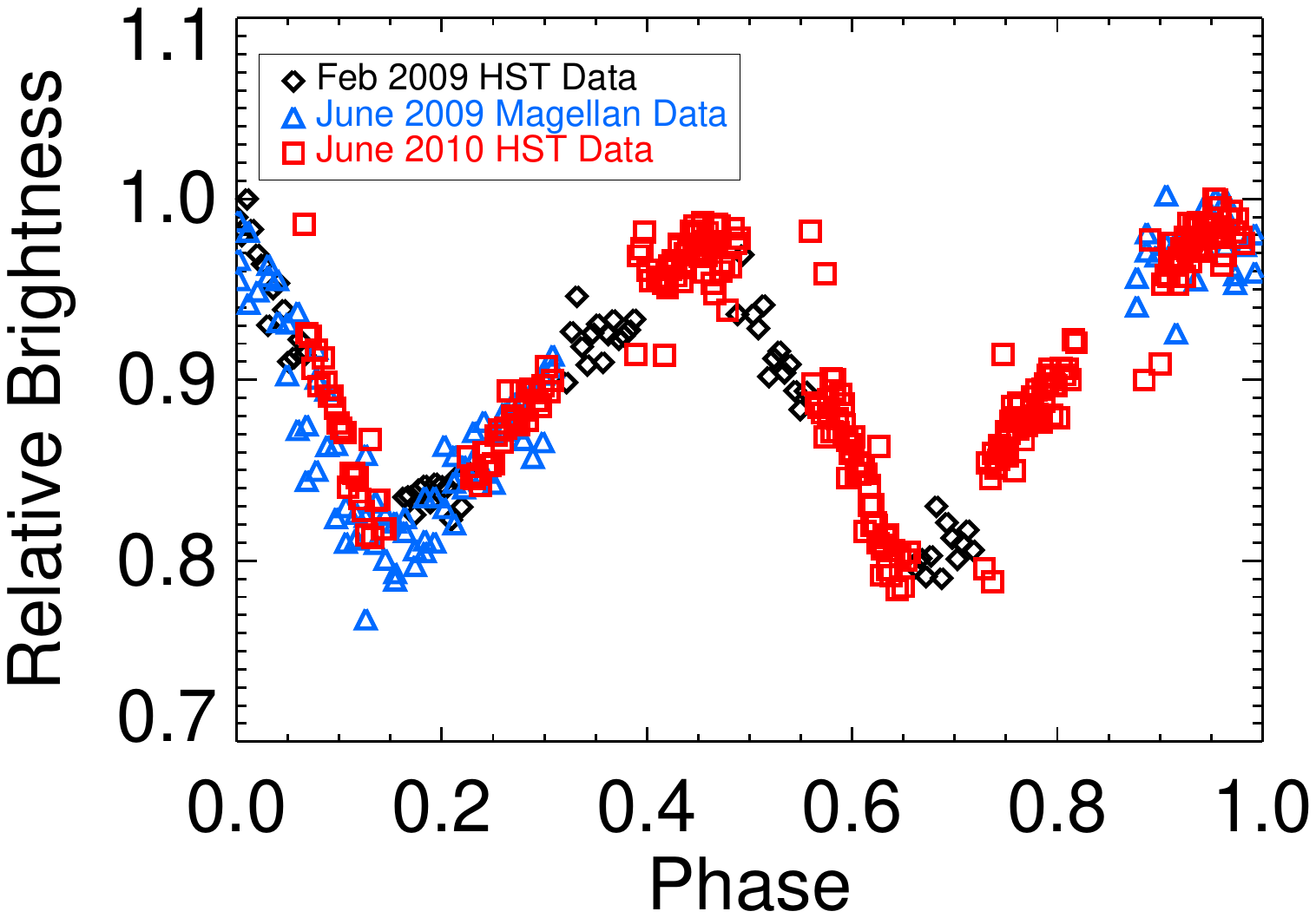}
    \caption{Phase folded light curve for Hi'iaka relative normalized photometry. Black diamonds correspond to HST data from February 4, 2009 (Program 11971); blue triangles correspond to Magellan Observations on June 1, 2009; and red squares correspond to HST data from June 28, 2010 (Program 12243). The data has been folded over a period of 9.79736 hours, but several plots with periods near 9.8 hours are similar. We conclude that the rotational period of Hi'iaka is approximately 9.8 hours with an amplitude of 19 $\pm$ 1\%.}
	\label{figure:phase_folded}
\end{figure}

The trial periods with the lowest reduced $\chi^2$ values were used to make a series of phase folded plots.  These were inspected by eye, as PDM2 only minimizes phase dispersion and does not invoke a smoothness criterion that is more consistent with a light curve. The phase folded plot that was determined to be best is shown in Figure \ref{figure:phase_folded}.  This plot corresponds to the trial period with the second lowest reduced $\chi^2$ value (9.79736 hours), but was considerably smoother than the plot for the lowest value (9.71141 hours). 

The PDM2 results show a forest of peaks corresponding to integer full rotations between our three disparate datasets. While additional work could potentially identify a more precise rotational period, this limited dataset establishes that Hi'iaka has an unexpectedly rapid rotation rate that is $\sim$120 times faster than the 49.5-day orbital period. 

\subsection{Implications for Hi'iaka's Shape}

From Figure \ref{figure:phase_folded}, we can see that the brightness variation is 19 $\pm$ 1\%, with a possible additional $\sim$1\% systematic error due to our assumptions in producing the normalized relative photometry. The sawtooth shape indicates an irregularly-shaped body, but without additional observations, we choose to approximate the shape of Hi'iaka as a tri-axial ellipsoid with semi-axes $a > b > c$. The three (degenerate) parameters that control the light-curve amplitude of a tri-axial ellipsoid are $b/a$, $c/a$, and $\theta$, the angle between the line-of-sight and the rotational pole. The relationship between these parameters and the amplitude of the brightness variations in magnitudes ($\Delta m$) is \citep{2013AJ....145..124B}: 
\begin{equation}
    \label{equation:delta_m}
    \Delta m=2.5\log\frac{a}{b}-1.25\log\left(\frac{a^2\cos^2\theta+c^2\sin^2\theta}{b^2\cos^2\theta+c^2\sin^2\theta}\right) .
\end{equation}
If Hi'iaka is nearly equator-on, then $\theta \approx 90^{\circ}$ which gives a maximum value of $b/a$ of approximately 0.81 for Hi'iaka. Another approximation to help break the degeneracy would be to limit $b/a$ and $c/a$ to values common for real solar system objects. We investigated $\Delta m$ as a function of $\theta$ for objects presumed to be of similar size to Hi'iaka (150 km radius), using Equation \ref{equation:delta_m}. The objects considered were Eugenia, Psyche, Camilla, Eunomia, and Hyperion.\footnote{These five objects were determined by investigating \texttt{ https://en.wikipedia.org/wiki/List\_of\_Solar\_System\_objects\_by\_size} for objects with volumetric radii near 150 km. When performing calculations based on these bodies, we assume they are represented by tri-axial ellipsoids with parameters as listed in this webpage. Thisbe, Phoebe, and Hektor also had similar sizes, but their shapes are very inconsistent with the observed Hi'iaka light curve and are not used. }  For these objects, $\theta$ between roughly 55-70 degrees results in the brightness change observed in the Hi'iaka light curve ($\Delta m \simeq 0.23$). Hi'iaka has certainly experienced a different formation and evolution environment than these objects, but if it is roughly similar in shape, then this would imply that perhaps $\theta \approx 60^{\circ}$ at the time of these observations. 

\subsection{Implications for Namaka's Light Curve}
Recall that the 2009 observations actually contain both Hi'iaka and Namaka in our aperture. (They are unresolved as the purpose of these observations was to detect a satellite-satellite mutual event.) We assumed that variability was due to Hi'iaka, which we now revisit.

Due to the near-commensurability between Hi'iaka's spin period and HST's orbital period (seen in Figure \ref{figure:lightcurve} for the 2010 data) and an unfortunate phasing, the 2009 and 2010 data provide almost exclusive coverage of Hi'iaka's phase curve. Therefore, it is not possible to rigorously compare the 2009 mutual event data with the ``true'' Hi'iaka light curve inferred from lightcurves at other epochs. The light curve is reasonably smooth, but this is partly by construction. Therefore, it is difficult to say for certain what effect Namaka's light curve or the possible mutual event had on the photometry.

Even if Namaka's light curve is entirely constant over this time interval, it would create a $\sim$20\% dilution of Hi'iaka's light curve, even in normalized photometry. This may be visible in Figure \ref{figure:phase_folded} near a phase of 0.2 where the February 2009 data are systematically brighter than the June 2009 Magellan data. Near 0.7 in phase, perhaps the true light curve of Hi'iaka is deeper than portrayed. In any case, the shape and structure of the light curve is preserved and our inference of a rotational period of 9.8 hours for Hi'iaka is not affected.

Inspection of Figure \ref{figure:phase_folded} suggests that major variability on short timescales beyond Hi'iaka's light curve is unlikely. Namaka is $\sim$4 times fainter than Hi'iaka in this filter, so a lack of variability at the $\sim$5\% level would suggest that Namaka is not more than $\sim$20\% variable. Given Namaka's size, it is likely to be aspherical like Hi'iaka (or more so). So, an apparent lack of Namaka's light curve would suggest either a slow rotation (much longer than 10 hours) or a face-on orientation (which would require significant obliquity, since Namaka's orbit is very nearly edge-on). We note that Namaka has not shown significant variability in other single-orbit HST data. The 2010 data was obtained near a Haumea-Namaka mutual event and no robust variability is detected, but Namaka is close or within Haumea's PSF, so this does not provide a strong constraint. Altogether, the data hints that Namaka's spin period is longer than roughly a day. Whether or not the data indicate a slowly rotating Namaka, it is worth noting that for orbital periods longer $\sim$1 day, Namaka's high eccentricity would likely result in a chaotic rotation due to spin-orbit resonance overlap \citep{1995Icar..118..181D,2000ssd..book.....M}. 

A Hi'iaka-Namaka mutual event (shadowing and/or occultation) would last up to 100 minutes and could result in a $\sim$25\% drop in flux. An event this strong is not detected. Grazing events that are shorter than about 30 minutes would be too weak to detect. In between is a wide range of possibilities but the combined light curve does not seem to contain any obvious mutual event. This does not entirely rule out a mutual event as it could have occurred during Earth occultation. The lack of a mutual event has weak implications for the possible orbits of Hi'iaka and Namaka which are beyond the scope of this work. 

\subsection{Implications for Haumea System Photometry}
Haumea has been the subject of significant photometric study, often without resolving Hi'iaka. Since Hi'iaka is $\sim$5\% as bright as Haumea and has a $\sim$20\% light curve, failing to account for Hi'iaka in unresolved photometry can introduce a $\sim$1\% error in understanding Haumea. As Haumea has a $\sim$25\% intrinsic variability \citep{2008AJ....136.1502R}, Hi'iaka's effect will only be important for precise measurements. 

Reviewing the observations for Haumea's ``Dark Red Spot'' \citep{2008AJ....135.1749L,2009AJ....137.3404L}, we do not believe that Hi'iaka's light curve has any effect on these conclusions, which are spread over multiple nights (and therefore would average out Hi'iaka's effect) and stronger than 1\%. 

On the other hand, \citet{2009AJ....137.4766R} predicted that Haumea and Namaka would undergo mutual events and several ground-based measurements were obtained by multiple teams. These mutual events are only a few percent in amplitude and Hi'iaka's light curve did provide some confusion. With the phase curve provided in Figure \ref{figure:phase_folded}, it should be possible to minimize the confusion from Hi'iaka's orbit, although this requires fitting Hi'iaka's spin phase until future work identifies Hi'iaka's spin period and phase more precisely. 

\subsection{Precession of Spin Axis}
\label{precession}

The process of satellite despinning is directly connected to the evolution of satellite obliquity (the angle between the satellite's spin vector and the primary-satellite orbit vector). As Hi'iaka is rapidly rotating, then there is a chance that it has retained a significant obliquity. Indeed, Pluto's small satellites have very high obliquities \citep[90-120 degrees][]{2016Sci...351.0030W} and a measurement of Hi'iaka's obliquity will similarly provide information on the formation and evolution of its spin. Hi'iaka's obliquity cannot be discerned in a single epoch, but Haumea's mass will cause Hi'iaka's spin vector to precess which would manifest itself as changes in the light curve shape over time. 

Without information about the shape and obliquity of Hi'iaka, we seek here to provide only an approximate sense of how precession affects the spin axis direction. We average over the ``fast angles'' that describe Hi'iaka's spin and orbital orientation and assume a circular orbit for simplicity. In this case, the obliquity ($\phi$) remains constant and the precessing angle is known as the equinox. The precession period is given by 
\begin{equation}
\label{equation:period}
P_{\rm Precession} = -\frac{2}{3} \frac{P_{\rm orb}^2}{P_{\rm spin}}\frac{C}{C-A}\frac{1}{\cos \phi}
\end{equation}
where C and A are the standard moments of inertia ($C > B > A$) and the spin, orbital, and equinox precession periods are labeled \citep[e.g.,][]{2008phea.book.....S}.  For a tri-axial ellipsoid of uniform density, the moments of inertia are
\begin{equation}\label{eqn:a}
    A=\frac{M}{5}(b^2+c^2)
\end{equation}
\begin{equation}\label{eqn:c}
    C=\frac{M}{5}(a^2+b^2)
\end{equation}
Since $C/(C-A)$ for Hi'iaka is unknown, the median value of the previously mentioned Hi'iaka-sized objects (2.34) is used for illustration.

The minimal precession period occurs when the obliquity is zero ($\phi=0$, i.e., alignment of the spin and orbit axes) which gives $P_{Precession} \simeq 26$ years. This first-order estimate indicates that Hi'iaka's spin precession could be visible within only a few years, as only a fraction of the precession cycle is required to provide observable changes in the light curve. 

Larger obliquities produce larger light curve changes, but also have slower precession periods. To illustrate this, we assume a simple linear precession of the equinoxes and calculate $\theta$ (the angle between the line-of-sight and Hi'iaka's spin axis) as a function of time. Without a more sophisticated model, we approximate Hi'iaka's orbit as fixed at epoch HJD 2454615.0 \citep{2009AJ....137.4766R} and do not include small ($\lesssim$3$^{\circ}$) changes in the orbital viewing angle. 

\begin{figure}
    \centering
    \includegraphics[scale=.75]{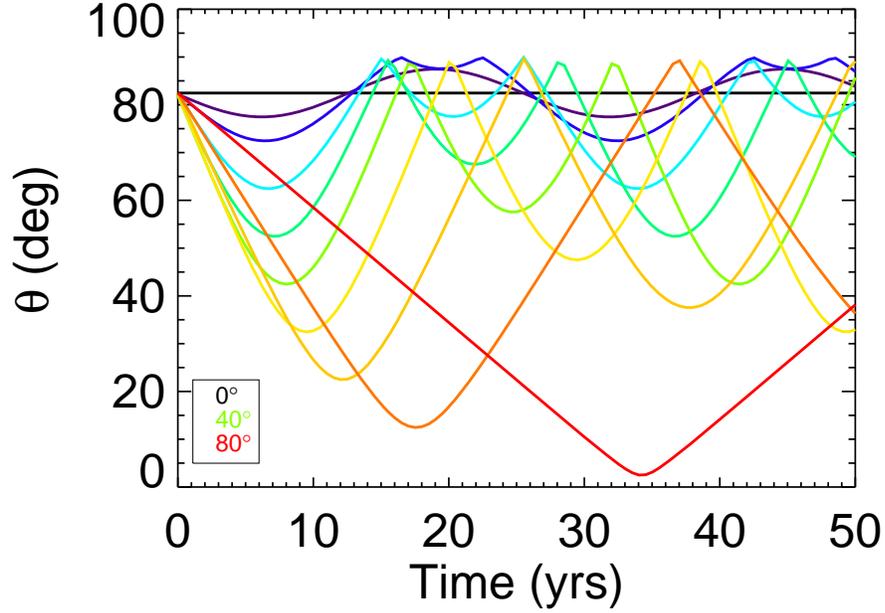}
    \caption{Plot of the angle between the Earth line of sight and the spin axis direction ($\theta$) over 50 years.  Each line is a different obliquity ($\phi$), starting at $0^{\circ}$ (black horizontal line; no precession), followed by $5^{\circ}$ (purple) and then $10^{\circ}$ (dark blue). The obliquity is then incremented by $10^{\circ}$ until $80^{\circ}$ (red). The y-axis ranges from $\theta = 0^{\circ}$ (sub-Earth point at Hi'iaka's pole) to $\theta = 90^{\circ}$ (sub-Earth point on Hi'iaka's equator), where values larger than 90 degrees are ``reflected'' since the effect on the light-curve amplitude (Equation \ref{equation:delta_m}) is symmetric. As shown in Equation \ref{equation:period}, larger obliquities are associated with longer precession periods. The light curve amplitude of Hi'iaka suggests the present value of $\theta$ around 60 degrees, with values less than 45 degrees unlikely. For clarity, the precessing equinox angle starts at the same value for each curve, but in actuality, the present day Hi'iaka could be anywhere in the precession cycle. Therefore, we cannot rule out any particular obliquity.} 
	\label{figure:precession_geometry}
\end{figure}

The results, which are only illustrative, are shown in Figure \ref{figure:precession_geometry}. For an obliquity of zero, Hi'iaka's spin is nearly equator-on (since the orbit is nearly edge-on at the fixed epoch). Increased obliquity leads to more pronounced changes, but also slower evolution, in accordance with Equation \ref{equation:period}. 

Based on our earlier sense from the shapes of Hi'iaka-sized objects, a present-day value of $\theta \approx 60^{\circ}$ would require an obliquity of $\gtrsim$20$^{\circ}$. Such a configuration suggests a precession rate of a few degrees per year. Zero obliquity is only possible if Hi'iaka is more spherical than other objects of similar size. By investigating objects with the full range of plausible shapes, only the most extreme objects could have $\theta \lesssim 45^{\circ}$ at the present epoch. This does not rule out any particular obliquity, since we do not know the phase of the equinox precession cycle. 

We find no significant evidence in our observations of a change in amplitude in the light curve due to the precession of the spin axis, but we present this as something to look for in future observations of the Haumea system. In the tri-axial approximation using shapes from objects similar in size to Hi'iaka, the light curve amplitude can change by $\sim$0.01 magnitudes per degree of change in $\theta$. Combined with a precession rate ranging from 1-10 degrees per year and that our observations are from $\sim$7 year old data, if Hi'iaka has a significant obliquity, it seems very likely that the light curve would be detectably different in new observations. New observations can confirm precession, but uniquely solving for the shape and obliquity of Hi'iaka would likely require sampling the light curve at multiple distinct epochs. In particular, the sawtooth shape of Hi'iaka's light curve (Figure \ref{figure:phase_folded}) indicates that a tri-axial ellipsoid is not a perfect model and Hi'iaka may be more angular. 

Our precession timescale estimates are sensitive to poorly known shape parameters, so the timescales could easily vary by tens of percents from what has been presented here. Still, observations of Hi'iaka spread over $\sim$10 years should be able to put valuable constraints on Hi'iaka's obliquity. 

For comparison, we also calculated the precession periods of the small moons of Pluto using Equation \ref{equation:period}, using dimensions and obliquity values assuming tri-axial ellipsoids with the parameters from \citet{2016Sci...351.0030W}. This yields precession periods of approximately 23, 3.0, 5.6, and 31 years for Styx, Nix, Kerberos, and Hydra respectively, though there is some uncertainty. The effect of Charon complicates their spin dynamics \citep{2015Natur.522...45S, 2015arXiv150606733C}, but hopefully long-term observations can provide information on these moons as well. 

\section{Formation Hypotheses}
\label{hypotheses}

The newfound result that Hi'iaka has a rapid rotation rate and potentially a significant obliquity helps provide insight to the formation and evolution of this moon, the Haumea system, and the Haumea collisional family, which we now explore in detail.

Hi'iaka is orbiting in a low eccentricity, low inclination orbit at $\sim$70 primary radii, which, combined with Namaka's similar orbital state, cannot be explained through known capture mechanisms \citep{2009AJ....137.4766R}. Thus, Hi'iaka formed around Haumea. There are two major end-member explanations for its present dynamical state: 1) Hi'iaka mostly formed near the Roche limit and dynamically evolved outward to its present location or 2) Hi'iaka mostly formed near/at its present location. 

At first glance, neither of these mechanisms is fully satisfying, even before considering the origin of Hi'iaka's rapid rotation. Some theories for the formation of the Moon include additional satellites \citep[e.g.,][]{1999AJ....117..603C} potentially long-lived \citep[e.g.,][]{2008phea.book.....S} and Haumea's satellites are in a broadly similar regime. However, dynamical evolution to its present location through tides seems to require extreme tidal parameters for Haumea. Using the volumetric radius, estimated physical parameters, and standard tidal equations requires unreasonable tidal parameters \citep{2013AJ....146...89C}. Recent work by \citet{2016arXiv160708591Q} shows that including the non-spherical nature of Haumea only gives a factor of 2 boost to the tidal evolution, overturning the argument of \citep{2009AJ....137.4766R} that this might be important. Although these first-order estimates fall far short,  
\citet{2016arXiv160708591Q} admits that additional analysis and an improved understanding of Haumea's size (which is still unknown), shape, and geophysical parameters may allow for such extensive tides to move from unrealistic to plausible. Even if strong tides can be invoked, for some values of the masses of the satellites, maintaining dynamical stability between interacting satellites with such significant tidal interactions is another major drawback to this hypothesis \citep{2013AJ....146...89C}. 

Formation at the present location avoids issues with tides, but prompts the question of why the proto-satellite disk extended to such a large semi-major axis, well beyond the regular satellite region of known bodies. Interactions with other objects (including Namaka) or the proto-satellite disk could have pushed Hi'iaka outward. In the case of Pluto, collisional expansion of the disk (or of multiple generations of satellites) can cause significant semi-major axis evolution \citep{2015AJ....150...11W,2015ApJ...809...88B}. These simulations included massive Charon, which is presumably a key component of the formation of Pluto's small moons, so these results are likely not relevant to the Haumea system. The most plausible hypothesis for such an extended disk around Haumea is that it formed subsequent to a collision onto a previous satellite (called the ur-satellite) of Haumea \citep{2009ApJ...700.1242S}. \citet{2013AJ....146...89C} explore this hypothesis in detail and find that it is mostly plausible. Further work is required to explain why this disk results in two widely-separated moons (at $\sim$35 and $\sim$70 primary radii), though once near this configuration, \citet{2013AJ....146...89C} suggests that resonant dynamics and standard tidal evolution can potentially reproduce the present eccentricities and inclinations. A downside to this hypothesis is that it removes any connection between the unusually rapid rotation of Haumea (caused by the initial impact) and the formation of a tight dynamical family, as the latter is independent of the former. \citet{2010ApJ...714.1789L} point out that SPH simulations cannot reproduce the creation of a rapidly-spinning primary and a relatively large ur-satellite, which suggests that this model may not be entirely self-consistent. Alternatively, since the formation of Haumea and the ur-satellite can occur early in the outer solar system when collisions are common, perhaps there are a reasonable set of collisions that form the ur-satellite and then spin-up Haumea (without destroying the binary). Even if forming a rapidly-rotating Haumea and an ur-satellite is reasonably probable, this formation hypothesis proposes that the spin-up event is effectively independent of the ur-satellite collision that forms the family. The combination of these two low probability events seems unreasonably low \citep{Campo-Bagatin13062016}; it is not clear why the only detectable collisional family in the Kuiper Belt would happen to form around the fastest spinning large body in the solar system. 

A full examination of these formation hypotheses is beyond the scope of this work, but we do investigate how Hi'iaka's rapid spin would fit into both of these end-member models. 

\subsection{Formation Close to Haumea and Evolved Out}
\label{close_in}

If Hi'iaka forms close to Haumea (near the Roche limit) and then evolves out, an initial expectation is that despinning tides would have slowed its rapid rotation early in its history when it was much closer to Haumea. 

The effect of despinning tides on Haumea can be parametrized in many ways. New models that explicitly include the expected frequency dependence of tides have been applied to some spin-orbit problems \citep[e.g.,][]{2009CeMDA.104..257E,2014ApJ...795....7M,2015CeMDA.122..359F}. Many second-order effects could be important such as solar interactions \citep[e.g.,][]{2012Icar..220..947P}, interactions with the other satellite\footnote{It is worth noting that the tidal dynamics of the Haumea system are unique among objects in the solar system. It has large dynamically interacting satellites like the giant planet satellite systems, but these evolve very slowly in semi-major axis due to weak tidal dissipation in gas and ice giants ($Q \gtrsim 10^4$). Among terrestrial/icy primaries with large dissipation, Haumea is unique in having two known moons which are both relatively massive and strongly interacting.} \citep{2013AJ....146...89C}, spin-orbit resonances and chaos \citep[e.g.,][]{1984Icar...58..137W,1995Icar..118..181D}, and  other potential issues. To simplify the problem into a tractable one and for comparison to previous work, we begin by using a simplified technique that can identify important dynamical results to approximately first order. Hence, we elect to use the ``classic'' constant Q models of \citet{1968ARA&A...6..287G}, keeping in mind that they are, at best, just approximations to a more complex history. 

In these models, the rate of change of the spin frequency $\omega$ is given by: 
\begin{equation} \label{omegadot}
\dot{\omega} = sign(\omega - n) \frac{3}{2}\frac{k_{2s}}{Q_s} \frac{1}{C_s} \left( \frac{M_p}{m_s} \right) \left( \frac{r_s}{a} \right)^3 \frac{GM_p}{a^3}
\end{equation}
where n is the mean motion, $k_2$ is the order two tidal Love number, Q is the tidal quality factor, M and m are masses, r is the radius, a is the semi-major axis, and G is the gravitational constant. In our case, the primary (``p'') is Haumea and the secondary (``s'') is Hi'iaka. 

\subsubsection{Initial Rough Quantitative Estimates}
\label{no_time}

In order to provide very rough quantitative estimates for the evolution of the Haumea-Hi'iaka system, we follow the method of \citet{2000ssd..book.....M} for estimating $k_2$ \citep[see also][]{2016arXiv160708591Q}:
\begin{equation} \label{equation:k2}
k_{2}=\frac{3}{2(1+\mu_{\rm eff})}
\end{equation}
where $\mu_{\rm eff}$ is the effective rigidity given by 
\begin{equation} \label{equation:rigidity}
\mu_{\rm eff}=\frac{19\mu}{2\rho g r}
\end{equation}
where $\mu=4\times10^9$ N~m$^{-2}$ is the assumed rigidity for an icy body , $\rho$ is the density, $r$ is the radius, and $g$ is the surface gravity.  We take $\mu$ from \citet{2000ssd..book.....M}, however the appropriate rigidity for tidal analyses could easily be off by orders of magnitude.

\begin{deluxetable}{llrrrrrrrrrc}
\tabletypesize{\tiny}
\tablecaption{Key Parameters for Planetary Satellites\label{table:parameters}}
\tablehead{\colhead{Object} & \colhead{Satellite} & \colhead{Mass} & 
\colhead{Radius\tablenotemark{a}} & \colhead{Density} & \colhead{$a$} & \colhead{$P_{orb}$} & \colhead{$P_{spin}$} & \colhead{$k_2$\tablenotemark{b}} & \colhead{$g$} & \colhead{$\tau_{\omega}$\tablenotemark{c}} & \colhead{Ref}\\
 & & (kg) & (km) & (kg~m$^{-3}$) & (km) & (days) & (days) & & (m~s$^{-2}$) & (years) &}

\startdata
Haumea & & $4.006\times10^{21}$ & $715$ & $2600$ & \nodata & \nodata & \nodata & $0.03$ & $0.3$ & \nodata &1,2\\
Haumea & Hi'iaka & $2\times10^{19}$ & $150$ & $1000$ & $49880$ & $49.462$ & $0.408$ & $0.0004$ & $0.06$ & $2\times10^{10}$ & 1,3\\
Haumea & Namaka & $2\times10^{18}$ & $75$ & $1000$ & $25657$ & $18.2783$ & \nodata & $0.00007$ & $0.02$ & $2\times10^{9}$ & 1,3\\
\hline
Pluto & & $1.304\times10^{22}$ & $1187$ & $1860$ & \nodata & \nodata & \nodata & $0.05$ & $0.6$ & \nodata & 4\\
Pluto & Charon & $1.59\times10^{21}$ & $606$ & $1700$ & $17540$ & $6.3872$ & $6.3872$ & $0.01$ & $0.3$ & $9\times10^5$ & 4\\
Pluto & Styx & $1.0\times10^{15}$ & $5.2$ & $1700$ & $42656$ & $20.1616$ & $3.24$ & $0.0000009$ & $0.002$ & $3\times10^{12}$ & 5,6\\
Pluto & Nix & $5.1\times10^{16}$ & $19$ & $1700$ & $48694$ & $24.8546$ & $1.829$ & $0.00001$ & $0.009$ & $4\times10^{11}$ & 5,6\\
Pluto & Kerberos & $1.5\times10^{15}$ & $6.0$ & $1700$ & $57783$ & $32.1676$ & $5.31$ & $0.000001$ & $0.003$ & $1\times10^{13}$ & 5,6\\
Pluto & Hydra & $6.5\times10^{16}$ & $21$ & $1700$ & $64738$ & $38.2018$ & $0.4295$ & $0.00001$ & $0.01$ & $2\times10^{12}$ & 5,6\\
\hline
Earth & & $5.9722\times10^{24}$ & $6371$ & $5515$ & \nodata & \nodata & \nodata & $1$ & $9.8$ & \nodata & 7\\
Earth & Moon & $7.3459\times10^{22}$ & $1738$ & $3341$ & $384400$ & $27.322$ & $27.322$ & $0.3$ & $2$ & $5\times10^{7}$ & 7\\
\hline
Eris & & $1.66\times10^{22}$ & $1163$ & $2500$ & \nodata & \nodata & \nodata & $0.09$ & $0.8$ & \nodata & 8,9\\
Eris & Dysnomia & $2\times10^{20}$ & $342$ & $1000$ & $37350$ & $15.774$ & \nodata & $0.001$ & $0.1$ & $2\times10^{8}$ & 8,10\\
\hline
Makemake & & $4.4\times10^{21}$ & $715$ & $2300$ & \nodata & \nodata & \nodata & $0.04$ & $0.6$ & \nodata & 11\\
Makemake & MK2 (4$\%$) & $2.8\times10^{18}$ & $87.5$ & $1000$ & $21000$ & $12.4$ & \nodata & $0.00008$ & $0.02$ & $2\times10^{9}$ & 12\\
\enddata
\tablecomments{Key parameters and results from tidal despinning calculations.}

\tablenotetext{a}{Radii for Haumea and the small satellites of Pluto are the volumetric radii, calculated using $R=\sqrt[3]{abc}$.}
\tablenotetext{b}{Value of $k_2$ calculated using Equation \ref{equation:k2}.}
\tablenotetext{c}{Despinning timescale in the current position of each satellite, calculated using Equation \ref{equation:tau_w} and $Q=100$.}

\tablerefs{
(1) \citet{2009AJ....137.4766R} \quad
(2) \citet{Lockwood2014} \quad
(3) \citet{2013AJ....146...89C} \quad
(4) \citet{2015Sci...350.1815S} \quad
(5) \citet{2015Natur.522...45S} \quad
(6) \citet{2016Sci...351.0030W} \quad
(7) \citet{2008phea.book.....S} \quad
(8) \citet{Brown1585} \quad
(9) \citet{2011epsc.conf..137S} \quad
(10) \citet{refId0} \quad
(11) \citet{2041-8205-767-1-L7} \quad
(12) \citet{2016arXiv160407461P} \quad
}
\end{deluxetable}

After assuming a value for the classic tidal dissipation parameter $Q$ and, implicitly choosing a frequency dependence and rheology for the body \citep{2009CeMDA.104..257E}, we can arrive at the classic estimate for the tidal despinning timescale:

\begin{equation} \label{equation:tau_w}
\tau_{\omega} = \frac{\omega}{\dot{\omega}} \approx \frac{2}{15\pi}\frac{Q_s}{k_{2s}}\left(\frac{\rho_s}{\rho_p}\right)^{3/2}\left(\frac{a}{R_p}\right)^{9/2}P_{orb}
\end{equation}

Similar calculations result in the timescale for changes in the orbital frequency $n$: 
\begin{equation} \label{equation:tau_n}
\tau_n=-\frac{2}{3}\tau_a=-\frac{2}{18\pi}\frac{Q_p}{k_{2p}}\frac{M_p}{m_s}\left(\frac{a}{R_p}\right)^5P_{orb}
\end{equation}
where $M_p$ is the mass of Haumea and $m_s$ is the mass of Hi'iaka.  

We find below that calling these ``timescales'' is inappropriate. However, starting with these equations results in estimates for tidal properties and timescales shown in Table \ref{table:parameters}. This initial rough analysis indicates that Hi'iaka's despinning timescale is longer than the age of the solar system. However, this conclusion requires many simplifying assumptions which we now explore in greater detail.

\subsubsection{Time Evolved Numerical Solution}
\label{time}

The standard equation for the despinning ``timescale'' ($\tau_\omega$) evaluated at the present location of Hi'iaka is a poor approximation of whether Hi'iaka could have despun. Under the assumption of active tidal evolution, the semi-major axis of the satellite has changed significantly. As the despinning tides are strongly dependent on $a$ ($\tau_{\omega} \propto a^6$), $\tau_{\omega}$ is certain to be an overestimate of the time required to despin a satellite. Furthermore, the initial spin period (which nominally determines the number of despinning timescales required for synchronization) is not known and could cover quite a range. 

To explore the actual tidal evolution, we developed a simple numerical model that calculates the time-evolution of the spin frequency during semi-major axis expansion. We continue to make the assumption of a simplified tidal model, no spin-orbit resonances or chaos, and neglecting outside influences (e.g., Namaka). 

Equation \ref{equation:tau_n} implies that the semi-major axis evolution has the form $a(t) = (a_f-a_0) (t/T)^{2/13} + a_0$, where $T$ is assumed to be the age of the solar system\footnote{In this hypothesis, where Hi'iaka is coeval with the Haumea family, we rely on the results of \citet{2009AJ....137.4766R} and \citet{2012Icar..221..106V} that the family is ancient with an age comparable to the age of the Solar System.}, $a_0$ is the initial semi-major axis, and $a_f$ is the present semi-major axis. 

At every timestep, the appropriate value of $a$ was then used to determine how the rate of change of the spin frequency, $\dot{\omega}$, changes with time. The spin is evolved following Euler's Method:  $\omega(t_i)=\omega_0+\Delta t\dot{\omega}(t_{i-1})$ where $\omega_0$ is the initial spin frequency, $i$ is the iteration number, $\Delta t$ is the time step, and $\dot{\omega}(t_{i-1})$ is the rate of change of the spin frequency for the previous iteration, calculated from Equation \ref{omegadot}. In order to resolve the evolution which is orders of magnitude more rapid at the beginning of the simulation, we use exponentially increasing values of $\Delta t$ so that the iteration timesteps $t_i$ are evenly spaced logarithmically. We checked many different values for the number of iterations and there were no issues with convergence. Under certain assumptions (always supersynchronous or subsynchronous, always reaching $a_f$ at time $T$, and retaining $a_0$ and $\omega_0$), the evolution equations can be solved for analytically and these results exactly confirmed the numerical simulations. 

\begin{figure}
    \centering
    \includegraphics[scale=.75]{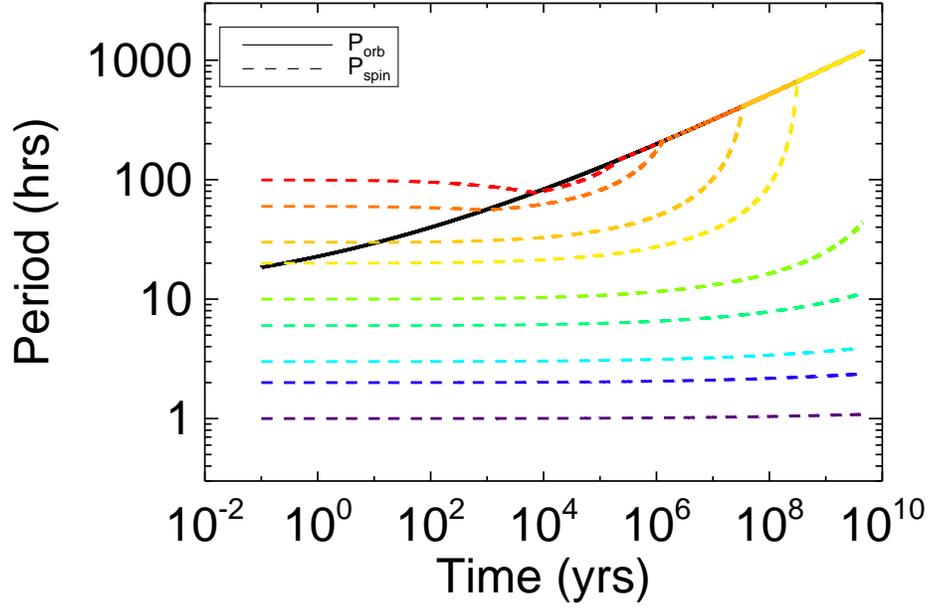}
    \caption{Results of the time evolved tidal despinning for Hi'iaka.  Physical and tidal parameters from Table \ref{table:parameters} were assumed along with an initial semi-major axis of $a_0=2000$ km were assumed.  Nine initial spin periods were tested (dashed lines) and compared to the evolution of the orbital period of Hi'iaka (solid line). Analytical solutions (not shown for clarity) match exactly those models. According to this tidal model, it is only possible for Hi'iaka to despin for longer initial spin periods. For initial spin periods comparable to the current value, the despinning cannot keep up with the orbital period and Hi'iaka never despins. A more detailed picture of the tidal parameters needed to despin Hi'iaka is shown in Figure \ref{figure:k2oqmod}.}
	\label{figure:time}
\end{figure}

Our nominal runs used the parameters given in Table \ref{table:parameters} and $Q_s = 100$. Multiple $a_0$ and $\omega_0$ values were tested and the results of one of these tests, with $a_0=2000$ km (just outside Haumea's Roche limit), is shown in Figure \ref{figure:time}. It is clear that the initial spin period is an important variable in determining whether or not Hi'iaka would be synchronous at the present time. Simulations with other values of $a_0$ are similar, but demonstrate that $a_0$ is also an important variable. 

We investigated whether it was possible to determine analytically the time needed to despin given actual semi-major axis evolution and retaining $a_0$ and $\omega_0$ in the solution (even though common practice is to neglect these). The result is a quartic polynomial in the despinning time to the 1/13 power with no path to a general solution. Furthermore, the analytical results were nearly as time-consuming to calculate as simply propagating the motion numerically. Hence, the numerical technique is preferred. 

For understanding how Hi'iaka's rapid spin rate affects our understanding of its formation, time evolution assuming certain parameters is only a first step. The more relevant question is: for what tidal parameters (e.g., $\frac{k_{2s}}{Q_s}$) does Hi'iaka despin in the age of the solar system, as a function of the initial semi-major axis and spin period? We combined our numerical technique with a bisection search in order to answer this question. Specifically, we calculate the value of $Z$, such that $\frac{k_{2s}}{Q_s} = Z \frac{0.0004}{100}$ leads to despinning within a few percent of the age of the solar system, where the numerical values are the nominal values listed in Table \ref{table:parameters}.\footnote{Equivalently, $Z$ could modify the unknown physical parameters, such as the radius of Hi'iaka.} The semi-major axis evolution of Hi'iaka starts at $a_0$ and ends at the present position in the age of the solar system, as before. 

Slight changes in the parameters ($a_0$, $w_0$, and $Z$) led to very large differences in the time needed to reach synchronous. This sensitivity is due to the strong dependence on semi-major axis, which is evolving rapidly. This is reflected in Figure \ref{figure:time} by the sharp down-turns in the computed spin evolution (even on a log-log plot). If synchroneity is just missed at an early epoch then it can take a very long time to ``catch up.'' 

\begin{figure}
    \centering
    \includegraphics[scale=.75]{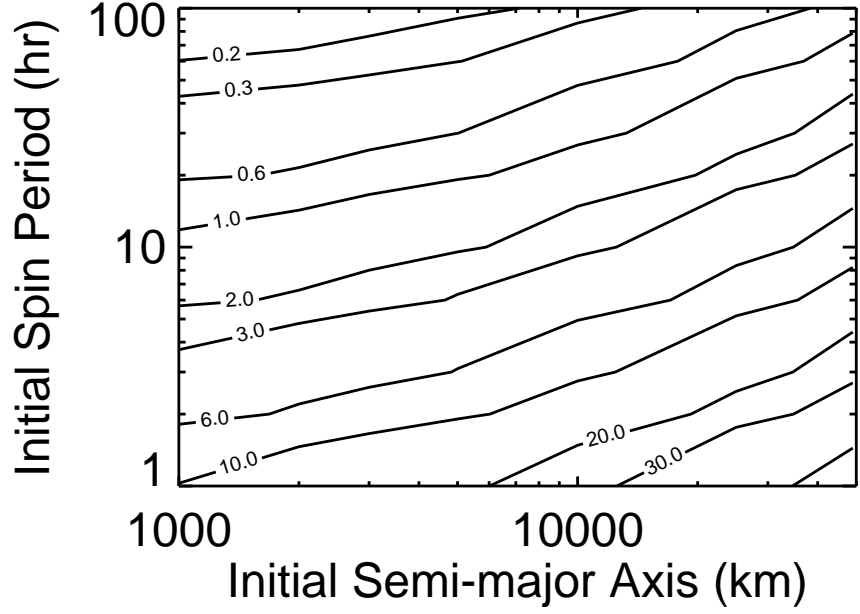}
    \caption{Contour plot of $Z$, the scale factor by which the nominal $\frac{k_{2s}}{Q_s} = 4 \times 10^{-6}$ must be multiplied in order for Hi'iaka to despin after $\sim$4.5 GYr under our assumed tidal model. We find that the initial semi-major axis and initial spin period are important with an approximate relation of $Z \propto a_0 \omega_0$. Uncertainty in tidal parameters allows for a wide range of plausible $Z$ values, making Hi'iaka's evolution unclear. If despinning tides are 100 times stronger than this estimate, then Hi'iaka can despin for any initial conditions including formation at its present location. If tidal despinning tides are 10 times weaker than the nominal estimate, then Hi'iaka would not despin for any initial conditions, including significant tidal evolution.}
	\label{figure:k2oqmod} 
\end{figure}

We show the value of $Z$ in a contour plot in Figure \ref{figure:k2oqmod}. Despite the strong dependence on parameters, $Z$ is approximately proportional to $a_0 w_0$. This is consistent with an analytical investigation of the dependence of $Z$ on these parameters. We have included $a_0 \simeq a_f$ as a prelude to the discussion below where Hi'iaka does not undergo tidal evolution. 

As the actual effective tidal parameters of Hi'iaka are uncertain by orders of magnitude, $Z$ is highly uncertain as well. These results show that, if $\frac{k_{2s}}{Q_s}$ is 100 times larger than the nominal value, then Hi'iaka will despin in practically any circumstance (no tidal evolution and initial very rapid rotation). If $\frac{k_{2s}}{Q_s}$ is 10 times smaller than nominal, then Hi'iaka would not despin in the age of the solar system, even if it started interior to the Roche lobe with a very slow spin rate. The wide variation in outcomes based on a relatively small uncertainty in tidal parameters is frustrating, but these results clearly indicate that Hi'iaka need not tidally despin. 

That a highly tidally evolved regular satellite could avoid despinning seemed contrary to our initial understanding. The common assumption for regular satellites is that, since the despinning timescale at the Roche limit is so small, the satellites quickly synchronize. And, once in a synchronous state, only ``small'' corrections are needed to maintain synchroneity as the satellite evolves outward due to tides. This then implies for a separate of timescales between the satellite synchronization and longer-term evolution of the semi-major axis that is often assumed \citep[e.g.,][]{1996Icar..122..166G,2009ApJ...691...54G}. Thus, such satellites are expected to be synchronous, even if the despinning timescale at the present position is longer than the age of the system.

We have identified significant issues with this common story. In particular, satellites of terrestrial planets experience significant tidal evolution (since $Q_p$ is so small, compared to gas giants). It is therefore plausible that semi-major axis evolution (or, more precisely, the evolution of the mean motion) is so fast that despinning tides simply cannot keep up. If satellite tides are weak enough compared to primary tides, there must be a regime where despinning cannot keep up with orbital expansion and the satellite does not remain synchronous. This is an arguable proposition even if our models for tides are incorrect compared to newer models based on appropriate geophysics. While our estimates of time-averaged approximate tidal parameters and corresponding timescales may be off by orders of magnitude, it remains the case that supersynchronous regular satellites like Hi'iaka could be explained by despinning rates that are slower than semi-major axis expansion. 

As an example of how despinning tides could be weak, we consider the recent analysis by \citet{2015AJ....150...98E} which proposes that the tidal dissipation rate of small bodies is controlled by viscosity ($\eta$), not strength ($\mu$). Using $\eta \approx 10^{15}$ Pa$\cdot$s for warm ice \citep{1989Icar...81..220O}, which is probably a strong (1-5 orders of magnitude) underestimate for the viscosity of Hi'iaka/Namaka, we find that Equation (65) in \citet{2015AJ....150...98E} shows that the dissipation rate $k_2/Q$ in these small bodies is in the regime where it is proportional to $1/(\eta \chi)$ where $\chi$ is the related to the spin or orbital frequencies ($\simeq 10^{-4}$). Using a quadrupole approximation ($l = 2$) and values from Table \ref{table:parameters} suggests that the effective $k_{2s}/Q_s$ in this geophysical model is $4 \times 10^{-6}$. This is 100 times weaker than our estimates based on the classical rigidity model above and would prevent Hi'iaka from despinning under any reasonable initial condition. Considering that cold\footnote{At present, the tidal dissipation in Hi'iaka is approximately 1 Watt for the whole body, indicating that tidal heating is not a significant source of heat; in a body so small, retained primordial heat would be minimal.} ice would have a much higher viscosity, the effective $k_{2s}/Q_s$ could potentially be as low as $10^{-9}$! These geophysical arguments would be sufficient to weaken tidal despinning of Hi'iaka to the point where it cannot keep up with synchronous and would maintain any rapid initial spin. 

Another issue with the common story is the assumption that satellites quickly despin since their despinning timescales are so short (e.g., due to formation near the Roche limit). This can ignore the also rapid semi-major axis evolution; if the despinning timescale changes from 100 years to 1000 years due to semi-major axis expansion that happens in 50 years, then synchroneity is not an inevitable outcome. Furthermore, the ``despinning timescale'' at any semi-major axis is not an effective way to estimate whether a satellite is despun, since it ignores the major influence of $a_0$ and $w_0$. Numerical simulations are more self-consistent. 

The idea that the satellite near-instantly evolves to a synchronous state also oversimplifies the effects of spin-orbit resonances and associated chaos. It is likely that small regular satellites experience chaotic spin evolution due to overlapping spin-orbit resonances as long as the spin period is within a factor of several times the (changing) orbital period \citep{1984Icar...58..137W,1995Icar..118..181D,1999ssd..book.....M}. While Hi'iaka is spinning much too fast for chaos now, when the orbital period was tens of hours it could have been in the chaotic regime and therefore not seriously affected by despinning tides. For despinning tides to succeed as satellites slow down, they must be able to pass through this chaotic barrier. Our results in Figures \ref{figure:time} and \ref{figure:k2oqmod} suggest that if the Hi'iaka \emph{ever} reaches a spin period as rapid as $\sim$10 hours, then despinning tides would not succeed at synchronization (with the nominal tidal parameters). A chaotic regime early in its evolution could have readily imparted such a rapid spin state.

Together, these results suggest that the common assumptions that imply inevitable synchronization of regular satellites do not hold up to more detailed investigation. In particular, Hi'iaka's rapid spin (either 9.8 or 4.9 hours) is not inconsistent with formation near the Roche limit followed by semi-major axis expansion to its present location. 

\subsubsection{Comparison to Other Systems}
\label{comparison}

One way of roughly validating our understanding of despinning is to check whether our model would match observations in other systems. For comparison, the numerical solutions described in Section \ref{time} were also tested on Namaka and the Earth-Moon system using the parameters given in Table \ref{table:parameters}. Though these methods do not include spin-orbit resonance or chaos and require major assumptions about tidal properties, the results of these tests are consistent with observations. The results for the other systems listed below follow from Table \ref{table:parameters} and Equation \ref{equation:tau_w}.
\begin{itemize}
    \item The Moon and Charon are able to despin, as expected.
    \item Namaka despins for reasonable initial spin periods (all tested initial periods except for 1 hr). 
    \item Styx, Nix, Kerberos, and Hydra do not despin in the age of the Solar System. 
    \item Dysnomia has likely despun. 
    \item Makemake's recently discovered moon, MK2, has likely despun. 
\end{itemize}
We note that the classification as ``despun'' really means that the current rotation rate is within the regime where low-order spin-orbit resonances are important. Understanding whether the moons reside in a resonance (not necessarily synchronous) or a regime afflicted by spin-orbit chaos will require additional observational and theoretical investigation. Furthermore, this assumes the nominal tidal parameters; as we saw for Hi'iaka, changes within the orders of magnitude uncertainty can lead to substantially different outcomes. 

Since the small satellites of Pluto are known to be supersynchronous like Hi'iaka, we discuss their results briefly here. The formation of these moons is not well understood. Although matching detailed simulations is problematic \citep[e.g.,][]{2008arXiv0802.2951L}, the near-resonant locations suggest that these moons may have been pushed outward during Charon's orbital evolution \citep[e.g.,][]{2006Sci...313.1107W}. As Charon is much larger than these moons, it evolves much more quickly, potentially reaching its current position in only $\sim$10 MYr \citep{1997plch.book..159D,2014Icar..233..242C}. Resonant expansion with Charon would have resulted in rapid semi-major axis expansion for these moons, orders of magnitude faster than their expected despinning timescales. As with Hi'iaka, even if these moons used to be much closer to Pluto, their semi-major axis expansion could have been so rapid as to stifle tidal despinning, leaving them with their observed rapid rotation rates and high obliquities.  

\subsubsection{Conclusions for the Tidal Evolution Hypothesis}

A detailed investigation into the tidal despinning hypothesis shows that regular satellites need not despin if they had moderately rapid initial spin rates and despinning tides that are weaker than rapid semi-major axis expansion. For reasonable parameters and classic tidal models, this is fully consistent with the rapid spin states observed for Hi'iaka and Pluto's moons. Hence, the supersynchronous rotation rate for Hi'iaka \emph{does not suggest that Hi'iaka was never close to Haumea}. Hi'iaka's spin rate does not weigh against the hypothesis that Haumea's satellites formed close to Haumea and experienced significant tidal expansion. 

Similar processes that control the spin rate of Hi'iaka also affect its obliquity on comparable timescales. Obliquities are also affected by the final tail of collisional formation and complex dynamics, such as Cassini states \citep[e.g.,][]{2007ApJ...665..754F}. Hence, we expect the same results to hold for Hi'iaka's obliquity: tides may not have affected it, even if it formed very near to Haumea. Whether, and how, Hi'iaka's obliquity affects our understanding of how it formed is beyond the scope of this work.

We note that the standard models for eccentricity tides have similar dependencies on semi-major axes as despinning tides. Therefore, the importance of including semi-major axis evolution  is also applicable to eccentricity tides. One additional complication is that both the primary and secondary contribute to eccentricity tides and often in opposite ways, as discussed by \citet{2013AJ....146...89C}. Depending on the tidal model, rapidly rotating Hi'iaka could actually result in eccentricity pumping even by the secondary \citep[e.g.,][]{MignardII}. Most of the concerns expressed above about inappropriate assumptions for despinning tides apply similarly to eccentricity tides. Yet, in most models, satellite despinning should occur more rapidly  than satellite circularization. The rapidly rotating Hi'iaka then could be strong evidence that Hi'iaka's eccentricity was not lowered due to satellite tides. As with the spin rate, resonances -- this time mean-motion resonances with Namaka -- preclude us from drawing conclusions about the initial state of Hi'iaka's orbit, as discussed extensively in \citet{2013AJ....146...89C}. 

\subsection{Formation Far Out}
\label{far_out}

In the hypothesis where Hi'iaka forms near its present location, different considerations are needed to understand its current spin state. Figure \ref{figure:k2oqmod} shows that, if Hi'iaka was always near its present location, it would only despin if the initial rotation period was long ($\gtrsim$100 hours) or the tidal parameters several times larger than the nominal value. Hi'iaka's current spin period could be comparable to its spin period after formation far from Haumea. In this sense, it is similar to irregular satellites that are also unlikely to despin \citep{2010Icar..209..786M}. 

Unfortunately, there is little detail about what we might expect for the initial spin period of Hi'iaka in the hypothesis where Hi'iaka forms far from Haumea. As the inclination of Hi'iaka, Namaka, and Haumea's equator are all highly consistent, this requires formation in a proto-satellite disk with damped inclinations and eccentricities \citep{2009ApJ...700.1242S, 2013AJ....146...89C}. An impact with an ur-satellite that creates Hi'iaka, Namaka, and the collisional family would initially create a huge cloud of debris that then participates in a collisional cascade which creates the low-inclination disk. At the frigid temperatures of the Kuiper Belt, the coagulation into satellites should follow entirely gas-free solid-body formation by accretion. In this case, the spin and obliquity of the final Hi'iaka is controlled by the last few stochastic collisions (see also Section \ref{haumea}). This suggests that the initial spin period and obliquity cannot be reasonably inferred. In particular, Hi'iaka's rapid spin is also consistent with the hypothesis that it formed near its present location. 

It is important to recognize that we have considered the end-member possibilities; a case where the satellites form far from the Roche limit but also experience significant semi-major axis evolution is also possible. In any of these models, despinning tides might somewhat slow an initially more rapid spin to the present 9.8-hour period. 

\subsection{Other Possible Spin Up Explanations}
\label{spin_up}

As shown in Figure \ref{figure:k2oqmod}, if despinning tides are $\sim$100 times stronger than the nominal estimate given in Table \ref{table:parameters}, then both formation hypotheses may predict in a near-synchronous rotation rate for Hi'iaka. Given the orders-of-magnitude uncertainty in tidal parameters, we also briefly consider other possible explanations for recently spinning up Hi'iaka. 

Based on Hi'iaka's physical and orbital properties, we can immediately rule out gravitational effects from the Sun (except perhaps as would be relevant for Cassini states) as well as radiation effects like Yarkovsky and YORP. Namaka is too small and too far away to exert a significant influence, except to contribute mildly to spin-orbit chaos. 

Haumea has the largest quadrupole moment of objects of its size and a rapid rotation rate. Hence, it is a candidate for considering whether some kind of spin-spin resonance was important \citep{2015ApJ...810..110B, 2015Natur.522...45S}. However, even in this extreme case, it seems unlikely that spin-spin resonances are a dominant effect on Hi'iaka \citep{2015ApJ...810..110B}. This is emphasized by the fact that Hi'iaka's spin period is 9.8 hours compared to Haumea's 3.9 hours. While potentially close to the 5:2 spin-spin resonance, it is very unlikely that there is an important dynamical influence from this weak resonance, particularly at the present distance of 70 primary radii. The apparent resonance could easily be due to the fact that any two periods will be coincidentally somewhat near some ratio of small integers. Future work that identifies a more precise spin rate of Hi'iaka can compare it to the known precise spin rate of Haumea \citep{Lockwood2014} to be sure. 

Another potential explanation for Hi'iaka's spin rate is a recent collision. Even if tides had despun Hi'iaka to a synchronous rotation rate, a collision can potentially reset the spin. The collision only needs to be as ``recent'' as a few despinning timescales at the present location of Hi'iaka (measured in GYr, but with significant uncertainty). Pluto's satellites \citep{2016Sci...351.0030W} show impact craters and certainly Hi'iaka is also subject to collisions. 

We do not consider explicitly the probability of any particular collision, but focus instead on identifying what kind of collisions are even plausible with the observed properties of Hi'iaka. The collision must provide a significant spin up without destroying Hi'iaka or significantly perturbing its near-circular orbit, which would gain a much higher eccentricity than observed (0.05) with a velocity change of only $\sim$10 m~s$^{-1}$. It turns out that this limits the range of plausible impactors, even when considering simple conservation of momentum and angular momentum. 

We considered several types of collisions which had a possibility of spinning Hi’iaka to its currently observed rotational rate. The two possible options were a small heliocentric impactor and a Haumea-centric satellite (now part of Hi'iaka). We considered two different scenarios for each case. For each scenario we created a simulation in MATLAB using Monte Carlo methods. A given simulation would randomly determine a number of parameters within reasonable ranges and, using conservation of linear and angular momentum, determine the result of a collision on Hi'iaka's spin and eccentricity. These simulations ignore a large host of known physical and geophysical effects of impacts, but their only goal is to identify whether there are any collisions that can possibly conserve momentum and angular momentum, spin up Hi'iaka, and leave it on a nearly circular orbit.

\subsubsection{Heliocentric Impactors}
\label{helio}

The heliocentric impactor case was tested using a "bullet" impactor with a mass between $10^{14}$ and $10^{16}$ kg, $0.001\%$ to $0.1\%$ the mass of Hi'iaka.  This impactor collided with a velocity between $300$ and $2500$ m~s$^{-1}$, consistent with heliocentric impactors given Haumea's orbit.  The first heliocentric scenario tested involved a cratering impact in which a uniform cone of material was ejected perpendicular from the impact direction (which may not be perpendicular to the surface of Hi'iaka). We found that in collisions which resulted in a rotational period of less than ten hours, the impactor was capable of imparting velocity changes of well over $1000$ m~s$^{-1}$, and it was impossible to impart any velocity kicks of less than $100$ m~s$^{-1}$. This magnitude of a velocity change would drastically change the orbit of Hi’iaka (orbiting at $\sim$75 m~s$^{-1}$), effectively ruling out this scenario.

The second heliocentric scenario involved one of the same impactors hitting with a very high impact parameter in a “hit and run” type collision, in which the impactor rebounds perpendicular to the impact location and continues on with some fraction of its original speed, imparting both a linear and angular kick to Hi’iaka. This type of collision was able to impart the observed spin rate with linear kicks of less than $10$ m~s$^{-1}$. However, these results were contingent on an impactor just barely clipping the surface of Hi’iaka and bouncing off at the same very low angle, usually leaving with only a quarter of its initial velocity. This type of collision is unlikely and physically, the bounce is improbable, so we conclude that this type of collision most likely was not the cause of Hi’iaka’s present state.

Although not surprising, we feel these results are sufficient to rule out heliocentric impactors as origins for Hi'iaka's spin. 

\subsubsection{Haumea System Impactors}
\label{haumea}

We also consider the possibility that Haumea had three satellites, one of which collided with Hi'iaka to spin it up. For the Haumea-centric impactor case, we considered two smaller satellites moving at lower speeds, with relative velocities at infinite separation ranging from zero (co-orbital) to $300$ m~s$^{-1}$. We first considered a merging event of the two small satellites. For purposes of modeling how the angular momentum is related to the final spin, the satellites were modeled as spheres, and the merging simply as the two spheres sticking together. The merging simulations were able to generate rapid spins while imparting linear kicks of less than 10 m~s$^{-1}$ which would preserve Hi'iaka's eccentricity. 

The final scenario involved a Haumea satellite, proto-Hi’iaka, more than half the mass of today’s Hi’iaka, being struck by a smaller satellite, a “rubble pile,” a loose collection of rock and ice held together by its own gravity. In this scenario, the smaller impactor collides with proto-Hi’iaka with some impact parameter, resulting in a shear in which part of the impactor is removed, joining proto-Hi’iaka, while the rest continues on unaffected. This simulation also yielded positive results, with a wealth of collisions which imparted the necessary spin with low linear kicks. 

While we did not evaluate the probability that Haumea-centric impactors yielding the present day Hi'iaka, the wide ranges of acceptable impacts suggest that this is a possible mechanism, though much more detailed simulations would be necessary to truly assess their plausibility. So, another hypothesis for Hi'iaka's spin is that it was despun, but a third satellite recently (within $\sim \tau_{\omega}$) collided with Hi'iaka to produce the observed spin. Given that the other hypotheses can also reproduce Hi'iaka's spin, Occam's razor would suggest that we need not invoke a previous third satellite.

These results also confirm that, wherever Hi'iaka formed, impacts with other Haumea-centric bodies in the formation disk could have readily provided a rapid spin while preserving its low eccentricity and inclination. 

\section{Conclusion} \label{conclusion}

In summary, our work has led to the following conclusions.
\begin{itemize}
\item Observations show that Hi'iaka has a clear light curve with a sawtooth shape and amplitude of 19 $\pm$ 1\%. The three datasets with sufficient information are consistent with a double-peaked rotation period of about 9.8 hours. Thus, Hi'iaka is rotating $\sim$120 times faster than its orbital period. 
\item Hi'iaka may also have a significant obliquity that would be imminently detectable as changes in light curve shape. 
\item Despinning tides do not necessarily produce synchronous regular satellites. The time needed to despin a satellite depends on the initial semi-major axis and rotation rate. Considering likely initial spin rates and rapid semi-major axis expansion allow for Hi'iaka to maintain its highly supersynchrnous rotation, even if it formed near the Roche limit for nominal tidal parameters. Therefore Hi'iaka's spin rate does not rule out significant tidal evolution.
\item Hi'iaka's spin rate is also consistent with a formation near its present location. 
\item Heliocentric impactors cannot spin up Hi'iaka without destroying it or severely affecting its orbit. However, Haumea-centric impactors can readily provide the observed spin. 
\end{itemize}

Unfortunately, Hi'iaka's spin does not provide a strong discriminator between different formation hypotheses, particularly given the large uncertainty in possible tidal parameters. Thus, we suggest that future work to identify better and more self-consistent models for the formation of the Haumea system and family should primarily focus on explaining other observations. 

\begin{acknowledgments}

We thank Marc Buie for sharing his IDL library, including the \texttt{PDM2} code that was used in this work. DR acknowledges discussions with Matija {\'C}uk and others concerning possible explanations for Hi'iaka's unexpectedly rapid spin rate. We wish to thank the anonymous reviewers for their comments and suggestions which improved the quality of this paper.

Support was provided by NASA through grant HST-GO-13873 from the Space Telescope Science Institute (STScI),  which  is  operated  by  the  Association  of  Universities  for  Research  in  Astronomy,  Inc., under NASA contract NAS 5-26555. 

\end{acknowledgments}

\software{ISIS \citep{2000A&AS..144..363A}, DAOPHOT II \citep{1538-3873-99-613-191, 1992ASPC...25..297S}, PDM2 (Marc Buie, personal communication)}


\end{document}